\documentclass
[aps,prl,twocolumn,showpacs,superscriptaddress,groupedaddress,epsf]{revtex4-1}%
\usepackage{graphicx}
\usepackage{dcolumn}
\usepackage{bm}
\usepackage{amssymb}
\usepackage{setspace}
\usepackage{amsmath}
\usepackage{amsfonts}
\usepackage{verbatim}%
\providecommand{\U}[1]{\protect\rule{.1in}{.1in}}
\begin{document}

\title{Casimir energy in the Gribov-Zwanziger approach to QCD}
\author{Fabrizio Canfora$^{1,2}$, Luigi Rosa$^{3,4}$\\$^{1}$\textit{Centro de Estudios Cient\'{\i}ficos (CECS), Casilla 1469,
Valdivia, Chile.}\\$^{2}$\textit{Universidad Andr\'{e}s Bello, Av. Rep\'{u}blica 440, Santiago,
Chile.}\\$^{3}$\textit{Dipartimento di Fisica, Universit\`a di Napoli Federico II ,
Complesso Universitario di Monte S. Angelo, Via Cintia Edificio 6, 80126
Napoli, Italia.} \\$^{4}$\textit{INFN, Sezione di Napoli, Complesso Universitario di Monte S.
Angelo, Via Cintia Edificio 6, 80126 Napoli, Italia.}\\{\small canfora@cecs.cl, rosa@na.infn.it} }
\date{\today}

\begin{abstract}
In this paper we study the Casimir energy of QCD within the Gribov-Zwanziger
approach. In this model non-perturbative effects of gauge copies are properly
taken into account. We show that the computation of the Casimir energy for the
MIT bag model within the (refined) Gribov-Zwanziger approach not only can give the correct sign but it also gives an estimate for the radius of the bag.

\end{abstract}
\maketitle


\section{Introduction}

The Casimir energy is a very important quantity in Quantum Field Theory since
it discloses relevant properties of the vacuum structure of the theory. In the
present paper, we will focus on a very important area of application:
Yang-Mills theory and the MIT-bag model (see, for instance,
\cite{chod74,MIT1,MIT2})

In the MIT-bag model, gluons and quarks are analyzed in a ball of finite
radius: the simple yet effective conjecture is that gluons and quarks are
confined at large distances but they are free on small enough scales. The idea
of the model is that a negative Casimir energy could confine the quarks within
the bag (the nucleon). Unfortunately, as it is well known since the finding of
Boyer \cite{boyer68}, the Casimir energy obtained with the usual perturbative
gluon propagator turns out to have the \textit{wrong sign} i.e. to be always
positive and only very recently it has been shown that it is possible to have
a negative Casimir energy in a bag or by using an infrared modified propagator
$\Delta\left(  k\right)  $ (which satisfies $\Delta\left(  0\right)  =0$)
arising from a numerical analysis of the Schwinger-Dyson equation
\cite{oxman05} or by using a surface impedance approach \cite{io2012}. Even if
it can be interpreted as a simple effective model, the MIT-bag model encodes
an obvious experimental truth. Namely gluons (as well as quarks) \textit{are}
confined within a bounded region and, correspondingly, \textit{there is} a
non-trivial Casimir energy associated to the gluons confined within the
bounded region. Therefore, \textit{the wrong sign} of the MIT Casimir energy
is not just the "failure of a simple effective model" since (whatever the
final explanation of confinement will be) \textit{there will always be a
Casimir energy associated to the gluons} which has the wrong sign. More
generally, in the theory of Casimir energy it is still an open issue to find
fundamental mechanisms able to generate a change of sign in the case of the ball.

Here we show that non-perturbative effects related to the elimination of the
gauge copies (predicted long ago by Gribov \cite{Gri78} see also
{\cite{Singer1}} \cite{Zw89}) can give, in fact, a Casimir energy with
negative sign and an attractive force. Since in correspondence of Landau gauge copies (we will work in
the Landau Gauge), the Faddeev-Popov operator has zero modes, their presence
could invalidate the path integral approach to non-Abelian Yang-Mills theory. The main advantages of the present approach
are that it allows both to identify clearly the physical origin of the Infra-Red
modifications of the gluon propagator (namely, the elimination of the gauge
copies) and to compute the corresponding Casimir energy analytically (due to the fact that the Gribov-Zwanziger propagator is known explicitly). 

Indeed, the (solution of the) Schwinger-Dyson equation gives in principle the
exact answer (see, in particular, \cite{PRLSD}) but due to its complexity
numerical analysis are always necessary. In particular, it is difficult to
extract the precise analytic form of the propagator together with the
corresponding singularities from a numerical fit. Hence, the understanding of
the\ analytic structure of the gluon correlation function in the complex
$p^{2}$-plane is\ a difficult task which is actually under intensive
investigation. We remind the reader to \cite{PRLSD,algo1,algo2,algo3} 
for an overiew of what is being done on this matter. For this reason, we will work within the Gribov-Zwanziger framework in the following. However, it is worth to point out that the requirement to have a well behaved Casimir energy in the Infra-Red is an important consistency check on the non-perturbative gluon propagator which can be used as a constraint on the analytic structure of the propagator arising from the Schwinger-Dyson equation as well. 
\\

In order to define the path integral in the presence of Gribov copies, a very
successful method is to restrict the path-integral to the region around
$A_{\mu}=0$, which is called Gribov region $\Omega$ (see, in particular,
\cite{Gri78,Zw82,Zw89,DZ89,Zwa96,Va92}), so that the Faddeev-Popov operator
$M^{ab}=-\partial D_{\mu}^{ab}$ is positive-definite: $\Omega=\{A_{\mu}%
^{a}|\partial_{\mu}A_{\mu}^{a}=0,M^{ab}>0\}$. In the case in which the
space-time metric is flat and the topology is trivial \cite{footnote1} this
approach coincides with usual perturbation theory when the gauge field
$A_{\mu}$ is close to the origin (with respect to a suitable functional norm
\cite{Va92}) and, at the same time, it takes into account the infra-red
effects related to the (partial) elimination of the Gribov copies
\cite{Zw89,MaggS,Gracey}.

In \cite{SoVar,SoVar2,SoVar3} \cite{soreprl} it has been shown that suitable
$d=2$ condensates allow an infrared behavior of the propagator in good
agreement with lattice data \cite{RGZlattice}. As it will be explained in the
following, the latter result also provides one with the first a priori
argument suggesting that the poles of the refined Gribov-Zwanziger gluon
propagator can be complex conjugates. The GZ model equipped with the
condensates is called the Refined Gribov Zwanziger model (RGZ). A detailed
review of RGZ can be found in \cite{nele}. Its action reads:
\begin{widetext}
\begin{eqnarray}
S_{RGZ}^{quad} &  =& \int d^{4}x\Bigl[  \frac{1}{4}F_{\mu\nu}^{2}+b^a\partial_\mu A^a_\mu+\bar{c}^a\partial_\mu  D^{ab}_\mu c^b+\bar{\varphi}^{ac}_\mu\partial_\nu  D^{ab}_\nu {\varphi}^{ac}_\mu-\bar{\omega}^{ac}_\mu\partial_\nu  D^{ab}_\nu {\omega}^{ac}_\mu+  \nonumber \\
& & g\gamma^2f^{abc}A^a_\mu\left( \bar{\varphi}^{ac}_\mu+
{\varphi}^{ac}_\mu \right)-g f^{abc}\partial_\mu \bar{\omega^{ae}_\nu} D^{bd}_\mu c^d+{\varphi}^{ce}_\nu+\gamma^4 d(N^2-1)\Bigr]  + \nonumber \\
&& \left\{ \frac{m^{2}}{2}A_{\mu}^{a}A_{\mu}^{a}%
-M^{2}\overline{\varphi}_{\nu}^{bc}\varphi_{\nu}^{bc}\right\} \label{eq:Srgz}
\end{eqnarray}
\end{widetext}where the first three terms within square brackets are the
standard Yang-Mills+gauge-fixing+ghost terms, $\varphi$ and $\omega$ are
ghosts field which implement the restriction to $\Omega$, the term
proportional to $\gamma^{4}$ comes from the horizon condition, $d=4$ is the
space-time dimension and $N$ is the number of colors. The terms in curly
brackets accounts for the condensate in the gluon sector and in the $\left\{
\varphi,\overline{\varphi},\omega,\bar{\omega}\right\}  $ sector.

The gluon propagator $\Delta_{RGZ}$\ corresponding to this action is:
\begin{equation}
\Delta_{RGZ}(k^{2})=\!\left(  \delta_{\mu\nu}-\frac{k_{\mu}k_{\nu}}{k^{2}%
}\right)  \!\frac{k^{2}+M^{2}}{k^{4}+(M^{2}+m^{2})k^{2}+\lambda^{4}}
\label{retrick8.5}%
\end{equation}
where the above analytic expression for the gluon propagator is valid (at least) at one-loop and, according to the lattice data \cite{DOV}, the values of the parameters
are $\lambda^{4}=2g^{2}N\gamma^{4}+m^{2}M^{2}\ $,\ $\ M^{2}+m^{2}%
\approx0,337\pm0.047\left(  GeV\right)  ^{2}$ and $\lambda^{4}\approx
0,26\pm0.005\left(  GeV\right)  ^{4}$. It is worth emphasizing that, at a
first glance, one could think that the above propagator $\Delta_{RGZ}(k^{2})$
in Eq. (\ref{retrick8.5}) (despite the excellent agreement with the lattice
glueballs spectra which is able to produce) must have some problem. The reason
is that the propagator $\Delta_{SD}(k^{2})$ which has been used in
\cite{oxman05} in order to get the correct sign in the computation of the
Casimir energy satisfies $\Delta_{SD}(0)=0$ while $\Delta_{RGZ}(0)\neq0$. In
fact, we will show here that the refined Gribov-Zwanziger propagator is indeed
able to produce the correct Casimir energy giving also a good estimate of the
radius of the bag. This is a strong consistency check of the possibility to
have complex conjugated poles since the present analysis shows that, with the
same propagator in Eq. (\ref{retrick8.5}), one can produce both the correct
Casimir energy and a very good glueball spectra.

All the subtle issues of regularization and renormalization of the Casimir
energy depend on the UV behavior of the propagator \cite{CasimirBook}. Both
the RGZ-propagator and the standard QCD propagator have the same UV behavior,
hence the refined Gribov-Zwanziger approach does not introduce new UV
divergences and it is possible to use the standard regularization techniques.

In \cite{iparticle} the authors showed that, using linear combinations of
$A_{\mu}^{a}$ and $\left\{  \varphi,\overline{\varphi}\right\}  $, it is
possible to diagonalize Eq.(\ref{eq:Srgz}) introducing the so called
\emph{i-particles}. Since we work at one loop it suffices to consider the
quadratic action only (in the following for the sake of simplicity we will
assume $M^{2}=m^{2}$):
\[
S_{Q}=\!\!\int\!d^{4}x\frac{\lambda_{\mu}^{a}\left(  m_{+}^{2}-\partial
^{2}\right)  \lambda_{\mu}^{a}}{2}+\frac{\eta_{\mu}^{a}\left(  m_{-}%
^{2}-\partial^{2}\right)  \eta_{\mu}^{a}}{2}\ ,\label{iparticle1}%
\]
where $m_{+}^{2}$ and $m_{-}^{2}$ are the complex conjugated poles in $k^{2}$
of ${\Delta}_{RGZ}$ in Eq. (\ref{retrick8.5}). The corresponding propagators
are (in the Landau gauge)%
\[
\left\langle \!\theta_{\mu\pm}^{a}\left(  -\partial^{2}+m_{\pm}^{2}\right)
\theta_{\mu\pm}^{a}\!\right\rangle =\left(  \!\delta_{\mu\nu}-\frac{k_{\mu
}k_{\nu}}{k^{2}}\!\right)  \frac{1}{k^{2}+m_{\pm}^{2}}\ \label{iparticle5}%
\]
with $\theta_{\mu\pm}^{a}=\{\lambda_{\mu}^{a},\eta_{\mu}^{a}\}$ respectively.

The refined Gribov-Zwanziger approach does not prevent the poles in $k^{2}$ of
${\Delta}_{RGZ}$ from being real (even if the actual lattice data
\cite{RGZlattice}\ tell us that the poles are, indeed, complex conjugated). We
will now show that the only way to get a negative Casimir energy in the
present framework is to have complex poles. Hence, the requirement to have a
consistent Casimir energy can be interpreted as the first \textit{a priori}
theoretical criterion which favors complex conjugated poles within the refined
Gribov-Zwanziger approach. The Casimir energy corresponding to these
i-particles can be computed using the standard techniques \cite{CasimirBook}.
At a first glance, one could wonder whether it will be real since the presence
of complex square masses would suggest a complex result. However, we will now
see that the same mechanism of cancellation of the complex poles disclosed in
\cite{iparticle} and \cite{iparticles2} is at work. In other words, the
Casimir energy is the simplest possible observable in which the complex poles
of the i-particles cancel and one gets a real physical result. The fact that
the gluon field is described in terms of particles with complex masses, is
considered in this context, a sign of confinement. Indeed for complex masses
there is no possibility of having a Kallen-Lehmann spectral representation of
the propagator with positive spectral function.

\section{Casimir Energy}

In the following, for the sake of brevity we will concentrate on the case of a
cubic surface of volume $a^{3}$. This is the situation most similar to the
spherical case and, at the same time, it allows for a relatively easy
presentation. In the conclusions we will give the results for the spherical
surface too but a detailed treatment shall be presented elsewhere
\cite{inpreparation}.

Using bag boundary conditions, the Casimir energy for each of the
$i-$particles (and for each color) can be written \cite{CasimirBook}:
\begin{equation}
E_{Cas\pm}= \frac{\pi}{2 a}\left\{  2\sum_{n,l,p=1}^{\infty}\omega_{nlp\pm
}+3\sum_{n,l=1}^{\infty}\omega_{0nl\pm}\right\}
\end{equation}
with $\omega_{nlp\pm}=\left[  n^{2}+l^{2}+p^{2}+(a m_{\pm})^{2}\right]
^{1/2}$. In the sum we have taken into account the fact that in this case only
one index at a time can be posed equal to zero otherwise the gauge potential
is zero. The sum must be regularized; to this extent we will consider a new
function $\omega_{nlp\pm}(s)=\left[  n^{2}+l^{2}+p^{2}+(a m_{\pm})^{2}\right]
^{-s}$ so that the final result is obtained in the limit $\omega_{nlp\pm}%
=\lim_{s\rightarrow-1/2}\omega_{nlp\pm}(s)$. To compute $\sum\omega_{nlp\pm
}(s)$ we use the binomial theorem: \begin{widetext}
\begin{equation}
\sum_{n,l,p=1}^{\infty}\omega_{nlp\pm}(s)   =
\sum_{k=0}^{\infty}\left(  \!\!%
\begin{array}
[c]{c}%
{s+k-1}\\
{s-1}%
\end{array}
\!\!\right)  (a m_{\pm})^{2k}(-1)^k\sum_{n,l,p=1}^{\infty}\left(  n^{2}+l^{2}%
+p^{2}\right)  ^{-k-s} ,
\end{equation}
\end{widetext}
the last sum can be computed by means of the so called Epstein-Zeta function
\cite{ambjorn,edery}%

\[
Z_{d}(s)={\sum_{n_{1}...n_{d},=-\infty}^{\infty}}\!\!\!\!\!\!\!\!^{\prime
}\left(  n_{1}^{2}+n_{2}^{2}+...+n_{d}^{2}\right)  ^{-s}%
\]
where the prime means that the $n_{i}=0$ case must be omitted. It results that
\cite{edery}
\begin{align}
P_{d}(s)  &  :=\sum_{n_{1}...n_{d},=1}^{\infty}\left(  n_{1}^{2}+n_{2}%
^{2}+...+n_{d}^{2}\right)  ^{-s}\nonumber\\
&  =\sum_{m=1}^{d}(-1)^{(m+d)}\left(  \!\!%
\begin{array}
[c]{c}%
d\\
m
\end{array}
\!\!\right)  Z_{m}(s)
\end{align}
where $Z_{m}(s)$ is the Riemann zeta function.

Thus
\begin{align}
E_{Cas}  &  =\frac{\pi}{2a}\sum_{k=0}^{\infty}\left(  \!\!%
\begin{array}
[c]{c}%
{s+k-1}\\
{s-1}%
\end{array}
\!\!\right)  \left[  (am_{+})^{2k}+(am_{-})^{2k}\right] \nonumber\\
&  \frac{(-1)^{k}}{2}\left\{  2P_{3}(k+s)+3P_{2}(k+s)\right\}  _{s=-\frac
{1}{2}}%
\end{align}
We observe that, if we assume $m_{\pm}$ to be complex conjugate, being the
Casimir energy the sum of two conjugate complex number, it will result real.
Defining $m_{\pm}^{2}=\rho^{2}e^{(\pm i\phi)}$ we find
\begin{align}
E_{Cas}  &  =\frac{\pi}{2a}\sum_{k=0}^{\infty}\left(  \!\!%
\begin{array}
[c]{c}%
{s+k-1}\\
{s-1}%
\end{array}
\!\!\right)  (-1)^{k}(a\rho)^{2k}\cos(k\phi)\nonumber\\
&  \left\{  2P_{3}(k+s)+3P_{2}(k+s)\right\}  _{s=-1/2}.
\end{align}
with $a\rho<1$ so to have a convergent series.
If $|m|=\rho=0$ the only contribution comes from $k=0$ and we recover the
usual value: $E_{Cas}=\frac{0.092}{a}$. At next order in $\rho^{2}$ we find
\begin{equation}
\frac{E_{Cas}}{E_{0}}=\Biggl(1-0.645(a\rho)^{2}C_{\phi}+1.819(a\rho
)^{4}C_{2\phi}\Biggr) \label{eq:cascubo}%
\end{equation}
with $E_{0}=\frac{0.092}{a}$ and $C_{\phi}=cos(\phi)$.

Note that in this equation we discarded the pole $\sim\frac{1}{s+1/2}$ terms
that must be subtracted during regularization as usual. A close look at
Eq.(\ref{retrick8.5}) shows that $\pi/2\leq\phi\leq\pi$ thus substituting the
values $\rho^{2}=0.510\pm0.005GeV^{2},\phi=1.908\pm0.046 rad$ obtained from
Eq.(\ref{retrick8.5}) we end up with
\begin{equation}
E_{Cas}=\frac{1}{a}\left(  0.092+0.010a^{2}-0.034a^{4}\right)
\end{equation}
which is negative for $a\geq1.341GeV^{-1}\sim0.26fm$. Quite interestingly we
observe that the energy is positive  if $a\leq0.26fm$ (i.e. more or
less a sphere of radius $0.13fm$) and negative  in the opposite
case. 

The analysis of the sphere is much more complicated: we give here the final
result only and refer to the more extensive treatment in \cite{inpreparation}%
:
\begin{equation}
\frac{E_{Cas}}{\tilde{E}_{0}}=\Biggl(1+1.711(a\rho)^{2}C_{\phi}+0.447(a\rho
)^{4}C_{2\phi}\Biggr) \label{eq:cassfera}%
\end{equation}
with $\tilde{E}_{0}=\frac{0.084}{a}$.

Thus
\begin{equation}
E_{Cas}=\frac{1}{a}\left(  0.084-0.024a^{2}-0.008a^{4}\right)
\end{equation}
and in this case we have a negative energy for $a\geq1.445GeV^{-1}\sim0.29fm$.

Looking at Eqs.(\ref{eq:cascubo}),(\ref{eq:cassfera}) it is easy convince
oneself that for real masses ($\phi=0$) there is no possibility of obtaining
negative values for the Casimir energy. On the contrary, assuming two complex
masses for the \emph{i-particles} allows for the possibility of having
negative Casimir Energy. In particular this is true if we take two complex
conjugate masses which is the scenario consistent with lattice data. We note
that assuming $M^{2}=m^{2}=0$, i.e. no condensate, pure Gribov-Zwanziger, we
still have the possibility of having negative Casimir energy. In this case,
indeed, Eq.(\ref{eq:cassfera}) become ($\rho^{2}=0.51,\phi=\pi/2$) so that:
\begin{equation}
\frac{E_{Cas}}{\tilde{E}_{0}}=\Biggl(1-0.030a^{4}\Biggr)
\end{equation}
and it is negative when $a\geq2.40$ (on the other hand, as it has been already
discussed, the glueballs spectrum favors the refined scenario). 

Having changed the analytical form of the energy a careful study of the behavior of the Casimir force $f_{Cas}=-\frac{d E_{Cas}}{da}$ is necessary, we have:
\begin{equation}
f_{cube}=\frac{1}{a^2}(-0.502 x^2 C_{2 \phi}+0.059 x C_{\phi}+0.092)
\end{equation}
with $x=(a\rho)^2$.
Studying the sign of the trinomial, we find that it is possible to have negative, i.e.
attractive, force  in the following range of the parameters:
$$0<\phi<\frac{\pi}{4}\cup\frac{3\pi}{4}<\phi<\pi \mbox{ and } x_c<x<1$$
with $x_c=\frac{1}{C_{2\phi}}
\left[  0.059C_{\phi}+0.273\sqrt{0.023+2.486C_{2\phi}}   \right]$.

The same analysis for the sphere gives:
\begin{equation}
f_{sphere}=\frac{1}{a^2}(-0.113 x^2 C_{2\phi}-0.144 x C_{\phi}+0.084).
\end{equation}
This time the force will be attractive for
$$0<\phi<\frac{\pi}{4} \mbox{ and } x_s<x<1$$
with $x_s=\frac{1}{C_{2\phi}}
\left[  -0.637C_{\phi}+0.031\sqrt{216+1007C_{2\phi}}   \right]
$.
Thus we observe that our estimates are compatible with the modulus square of the gluon condensate $\rho^2=0.51$ with a bag radius $0.92<a<1.40$ but not with the phases. We note, also, that it is possible to have attractive force with real gluon condensate masses but  not without gluon condensate which corresponds to
$\phi=\pi/2$.

We note also
that, with $M^{2}\neq0$ the propagator is not vanishing when $k\rightarrow0$.
Thus, within this context, the attractive or repulsive nature of the Casimir
force does not depends just on the behavior of the propagator for $k^{2}=0$
but also on his detailed analytic form. For this reason the present results
are consistent with the ones in \cite{oxman05}.

In a sense, this paper is the complement of \cite{oxman05}. They find an
attractive Casimir force for the MIT-bag model by means of a modified
propagator of the Schwinger-Dyson type, while our result is obtained in a self
consistent (from the field theory point of view) scenario: the refined
Gribov-Zwanziger model of QCD. Both results show that non-perturbative
infra-red effects may give rise to quite substantial changes in the behavior
of the Casimir force.
In one approach, the RGZ gluon propagator {\it  with complex conjugate poles} reproduces  in an accurate way the lattice data,  while in the Schwinger Dyson approach recent numerical studies  point at  a structure with no complex poles  \cite{PRLSD} in the $p^2$-plane. 
As it has been already emphasized, the understanding of the analytic structure of the gluon correlation function in the complex $p^2$-plane is a difficult task which is actually under  investigation \cite{PRLSD,algo1,algo2,algo3}.  Only experiments will give us the exact form of the gluon propagator.

Some final  comments on the ghosts' contribution are in order. It is clear that
our computation concerns gluons only: we worked in a pure Yang-Mills theory.
In the usual situation, because of the Slavnov-Taylor identity the
contribution coming from the ghosts fields compensates with the one coming
from the gauge fixing leaving a term proportional to the ghost mass only
\cite{christensen}. Thus, in this case, assuming that the ghost masses are
very small, the contribution is negligible. Unfortunately, this time BRST
symmetry is broken (even though softly \cite{SoVar2}), and not all the Ward
identities are satisfied. Hence, we cannot be sure that the cancellation
mechanism still works and, to make an estimate, a complete knowledge both of
the ghost spectrum and of the corresponding boundary conditions, which is
still lacking, is necessary. 
However, it is possible to give the following
bound on their contribution (which we will call $E_{rest}$) to the Casimir
energy which is valid at one-loop order (consistently with the expression for the gluon propagator in Eq. (\ref{retrick8.5})). The one loop ghost propagator has the form \cite{SoVar3}:
$G^{ab}(k^{2})=\delta^{ab}\frac{1}{k^{2}}(1+\sigma(k^{2}))$ with $\sigma
(k^{2})\leq1$ because of the no pole condition \cite{SoVar2}. Thus the
\textit{worst situation} is the one in which the ghosts contribute to the
Casimir energy like two massless vector fields (but it is worth emphasizing
that ghosts usually contribute in the opposite way):
\[
\left\vert E_{rest}\right\vert \leq\frac{0.168}{a},
\]
which shows that, even including them, the conclusion of the paper on the
change of sign of the Casimir energy still holds with $a\simeq2.08$.
This very promising result (which is valid at one-loop within the (refined) Gribov-Zwanziger approach) suggests that, when the sum of all the contributions of the single terms in Eq.(\ref{eq:Srgz}) is properly taken into account, the Casimir energy, and correspondingly the Casimir force,
will have the correct sign as well.

\section{Acknowledgments}

The authors warmly thank Bruno Werneck Mintz for pointing out the most transparent way to interpret the results. This work is supported by Fondecyt grant 1120352. The Centro de Estudios Cient%
\'{}%
\i ficos (CECS) is funded by the Chilean Government through the Centers of
Excellence Base Financing Program of Conicyt. F. C. is also supported by
PROYECTO INSERCI\'{O}N CONICYT 79090034 and by the Agenzia Spaziale Italiana (ASI).


\begin{thebibliography}{99}
\bibitem {Singer1}Singer \cite{singer} showed that Gribov ambiguities occur in
all the gauge fixing conditions involving derivatives (see also \cite{jackiw} )
\bibitem {footnote1}{When the space-time metric and/or the topology are not
trivial, the pattern of appearance of gauge copies can change completely:
\cite{new1,new2,new4}.}
\end{thebibliography}
\end{document}